\documentclass[reprint, amsmath,amssymb, aps, showpacs, superscriptaddress, prx]{revtex4-1}

\usepackage{graphicx}
\usepackage{dcolumn}
\usepackage{bm}
\usepackage{mathtools}
\usepackage{amsfonts}
\usepackage{amsmath}

\def\expect#1{\left\langle #1 \right\rangle}
\def\bra#1{{\langle #1 |}}
\def\ket#1{{| #1 \rangle}}

\begin{document}
\preprint{}

\title{Universal quantum computation by the unitary control of ancilla qubits and using a fixed ancilla-register interaction}
\author{Timothy. J. Proctor}
 \email{py08tjp@leeds.ac.uk}
\affiliation{School of Physics and Astronomy, E C Stoner Building, University of Leeds, Leeds, LS2 9JT, UK}
\author{Erika Andersson}
\affiliation{SUPA, Institute for Photonics and Quantum Sciences, Heriot-Watt University, Edinburgh EH14 4AS, UK}
\author{Viv Kendon}
 \email{V.Kendon@leeds.ac.uk}
\affiliation{School of Physics and Astronomy, E C Stoner Building, University of Leeds, Leeds, LS2 9JT, UK}
\date{\today}

\begin{abstract}
We characterise a model of universal quantum computation where the register (computational) qubits are controlled by ancillary qubits, using only a single \emph{fixed} interaction between register and ancillary qubits. No additional access is required to the computational register and the dynamics of both the register and ancilla are unitary. This scheme is inspired by the measurement-based ancilla-driven quantum computation of Anders \emph{et al.} [PRA 82, 020301(R), 2010], but does not require measurements of the ancillas, and in this respect is similar to the original gate based model of quantum computation. We consider what possible forms this ancilla-register interaction can take, with a proof that the interaction is necessarily locally equivalent to SWAP combined with an entangling controlled gate. We further show which Hamiltonians can create such interactions and discuss two examples; the two-qubit XY Hamiltonian and a particular case of the XXZ Hamiltonian. We then give an example of a simple, finite and fault tolerant gate set for universal quantum computation in this model.
\end{abstract}


\maketitle

\section{Introduction}
Quantum computing promises fundamentally faster computation than classical computers could ever provide \cite{Feynmann85}, but the experimental challenges in building a quantum computer are formidable.  The basic theoretical setting for quantum computation, the quantum circuit model \cite{Feynmann85,Barenco95}, has been studied in detail for many years.  To implement a quantum circuit directly requires accurately controlled unitary dynamics in the form of one and multi-qubit gates on a computational register of qubits.  However, both the individual qubit addressability required for single-qubit unitaries and the multi-qubit entangling interactions can be experimentally challenging, each requiring different conditions to optimise their performance.  Other schemes have therefore been developed. One such scheme is measurement-based computation, or the one-way quantum computer, introduced by Raussendorf \emph{et al.} \cite{Raussendorf01,Raussendorf03}. In this scheme, the computation is achieved entirely by single-qubit measurements applied to a highly entangled initial state and classical feedforward dependant on the outcome of each measurement \cite{Briegel01,Hein04,Schlingemann04}. This separates the process into two steps; first the creation of entanglement which can be done by global operations, and then accurate single-qubit measurements. Another way to increase the physical viability of a scheme is to make use of `always on' interactions, such as the Heisenberg Hamiltonian for a spin chain: this allows the computation to be controlled entirely by precisely-timed global operations \cite{Benjamin03,Hu07,Benjamin04,Benjamin204,Lloyd93}.
\newline
\indent
Alternatively, direct control of the computational qubits can be avoided altogether by 
using ancilla systems to mediate the interactions.  One such scheme, based on the gate 
model of quantum computation, is the quantum bus model, or qubus. The qubus employs a continuous variable ancilla to mediate interactions between qubits in a computational register. This is done either with homodyne detection of the bus mode \cite{Munro05,Nemoto04} to complete the gates, or via entirely unitary dynamics \cite{Spiller06,Proctor12,Brown11}.  In the latter case case, either some access to the qubits for local operations, or more than one form of bus-register interaction, is required. Although only one interaction Hamiltonian is necessary, in general the interaction time must be varied to produce different gates.
\newline
\indent
Quantum computations are fragile to the effects of noise, therefore error correction schemes are required, to build in fault-tolerance. Error correction works by duplicating the information in one logical qubit across several physical qubits by forming an entangled state, which can then be used collectively to detect and correct uncorrelated errors arising from noise or inaccurate control operations \cite{Shor95,Steane96,Knill98}.  Examples of such schemes include those based on a lattice of qubits with only nearest neighbour interactions \cite{Raussendorf07,Fowler12,Fowler11,Fowler13}, which can be used to develop fault tolerant fully scalable architectures for quantum computers \cite{Jones12}. These error correcting schemes can be layered on top of any universal physical qubit substrate, so long as the architecture avoids the propagation of correlated errors, which are problematic to reliably detect or correct \cite{Klesse05}.  This motivates the study of simple physical quantum 
architectures which can be fully characterised against the requirements of fault-tolerance and error correction.
\newline
\indent
A hybrid between the circuit model and measurement-based computation, known as ancilla-driven quantum computation (ADQC), was recently developed by Anders \emph{et al.} \cite{Anders10,Kashefi09}. Like the qubus, no interactions are needed between register qubits.  The computation can be performed using a single fixed interaction between the ancilla and a register qubit, with full control of the ancillary qubit. The unitary evolution of the register is then driven by back-action onto the register, from measurements of the ancilla. The evolution is deterministic up to corrections that can be accounted for in the feed-forward, as in the one-way quantum computer \cite{Raussendorf01,Anders10}. In both this model and the one-way computer, for a fixed inaccuracy of measurement, the gate fidelity decreases with increased overall entanglement \cite{Morimae110,Morimae210}.
\newline
\indent
In the work presented here, we take a route to universal quantum computation similar in certain respects to both ancilla-driven and qubus computation. We begin, as with the ancilla-driven model, with the physically motived constraint of a register of computational qubits to which no access is allowed except through a single interaction, $K \in U(4)$, between an ancilla qubit and a single register qubit at a time. This is relevant to many experimental setups, particular where there are low-decoherence qubits between which it is hard to implement interactions. If it is possible to engineer an interaction with an ancillary system over which there is greater control, but which could have a shorter decoherence time, schemes such as the one presented here may be employed. Physical systems with such properties include nitrogen-vacancy (NV) centres in diamonds, where a nuclear spin is strongly coupled to the electron spin of the NV centre \cite{Shim13,Childress06,Dutt07}, and the coupling of spin qubits via flying photonic qubits \cite{Carter13}. It should however be noted that, in both the model presented here and in ADQC, the computational qubit data is stored briefly in the ancillary qubits and so is vulnerable to the decoherence rate of the ancillary system for short periods of time. The total time the data spends in the ancillary system will scale linearly with the number of gates applied.
\newline
\indent
Looking for a simple, fully unitary model, we enforce further constraints on our scheme. We will develop a model whereby the ancilla qubit is fully controllable, but where no measurements (except for final readout) are required. By fully controllable we mean that we may perform single-qubit unitaries on the ancilla. Although in practice multiple ancillas would be employed to implement gates in parallel, we will add the restriction that no ancilla-ancilla interactions are required since these could propagate correlated errors. Furthermore we wish to minimise the required physical interactions that need to be engineered, and in any physical realisation of such a scheme it is highly likely that different complementary systems will be employed as ancillary and register qubits. Hence any additional interaction between the ancilla will require a further physically distinct interaction to be engineered. 
\newline
\indent
For universal quantum computation, we need to be able to implement, via the ancilla, both single-register-qubit gates and multi-register-qubit entangling gates. It is desirable to minimise the number of interactions required for each gate. We will see that any single-qubit gate on a register qubit can be implemented using only two interactions with the ancilla, with unitary control of the ancilla applied in between. This is in contrast to ADQC, where one interaction is sufficient. Although for universal computation only a finite set of single-qubit gates is required, and in practice this is all that will be used, we will determine the interactions that allow \emph{any} single-qubit gate to be performed. This is so that the interactions can be considered universal for our model of computation regardless of which unitaries can be implemented on the ancilla (as long as some finite universal set can be implemented).
\newline
\indent
It is not possible to implement entangling gates between register qubits via an ancilla sequentially with only unitary dynamics \cite{Lamata08}, that is, by interacting an ancilla with one qubit and then the other without interacting with the first qubit again. This is in contrast to ADQC, where the use of measurement enables this. We will show, however, that we can perform two-qubit gates with only three ancilla interactions, two with one qubit and one with the other. We will begin by giving a simple example of an interaction with which we can perform universal quantum computation within the strong constraints of this model, and from there we will characterise \emph{all} possible forms this interaction could take. We will then discuss possible Hamiltonians with which this could be implemented and in doing so give a suitable finite universal gate set. We will refer to our model throughout as ancilla-controlled quantum computation (ACQC).

\section{An interaction for ACQC}
In what follows, all two-qubit unitaries will be denoted by upper case roman letters and single-qubit unitaries by lower case roman letters, with exceptions for well-known operators, such as the Pauli operators ($I$, $X$, $Y$, $Z$) and the Hadamard gate ($H$), where the appropriate standard notation is used.
\newline
\indent
Define the computational basis, $\{\ket{0},\ket{1}\}$, by the $+1$ and $-1$ eigenstates of the Pauli $Z$ operator respectively. Using the standard definitions, let 
\begin{equation} \text{SWAP}=\ket{00}\bra{00} + \ket{01}\bra{10} + \ket{10}\bra{01} +\ket{11}\bra{11}, \end{equation} 
\begin{equation} C(Z)= \ket{00}\bra{00} + \ket{01}\bra{01} + \ket{10}\bra{10} - \ket{11} \bra{11}, \end{equation} 
which can also be written as matrices in the computational basis:
\begin{equation} \text{SWAP} = \begin{pmatrix} 1 & 0 & 0 & 0 \\ 0 & 0 & 1 & 0 \\ 0 & 1 & 0 & 0 \\ 0 & 0 & 0 & 1 \end{pmatrix}, \end{equation}\begin{equation} C(Z) = \begin{pmatrix} 1 & 0 & 0 & 0 \\ 0 & 1 & 0 & 0 \\ 0 & 0 & 1 & 0 \\ 0 & 0 & 0 & -1 \end{pmatrix}. \end{equation}
\indent
If we take the ancilla-register interaction of
 SWAP, it is trivial to perform the single-qubit unitaries, simply by swapping the state of the register qubit and ancilla, performing the unitary on the ancilla and then swapping the states back. However this interaction is not entangling, so it cannot create entanglement between register qubits. To use the SWAP interaction we would need a second ancilla-register interaction, or an ancilla-ancilla interaction, both of which are forbidden in this model. Instead we can use the same method for the single-qubit gates using a different interaction. Let the fixed interaction be of the form $K=SC(Z)$ where 
\begin{equation} SV \equiv \text{SWAP} \cdot V,\end{equation}
$V\in U(4)$. Then 
\begin{equation} K = \ket{00}\bra{00} + \ket{01}\bra{10} + \ket{10}\bra{01} - \ket{11}\bra{11}, \end{equation} which in matrix notation is given by
\begin{equation} K = \begin{pmatrix} 1 & 0 & 0 & 0 \\ 0 & 0 & 1 & 0 \\ 0 & 1 & 0 & 0 \\ 0 & 0 & 0 & -1 \end{pmatrix}. \end{equation}
We will now show that ACQC is possible with this interaction, provided the ancilla is initialised in the $\ket{0}$ state. We first of all show how to perform arbitrary single-qubit unitaries on a register qubit (subscript $R$), by interacting the register qubit twice with the ancilla (subscript $A$). By linearity, we do not need to consider the whole register. It is sufficient to consider the action that this operation has on a general single register qubit state, $\ket{\phi}$, and an ancilla qubit initially in the state $\ket{0}$.
\begin{figure}[h!]
\center
\includegraphics[scale=1.2]{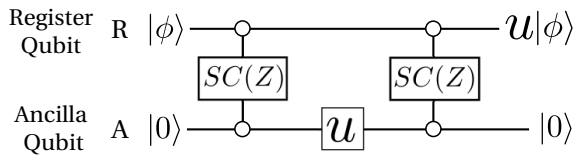}
\caption{\label{singlequbitfig} A simple example of a single-qubit gate in ACQC where the interaction is $SC(Z)$.}
\end{figure}
\newline
\indent
 As our ancilla is initialised in the state $\ket{0}$, the initial state is $\ket{0} \ket{\phi} \equiv \ket{0}_A \otimes \ket{\phi}_R$, where $\ket{\phi}$ is an arbitrary state of one qubit. We will omit the subscripts to denote which qubit(s) a state represents, or an operator acts on, when no ambiguity will arise. We first interact the register qubit and the ancilla: it is simple to confirm that $K \ket{0} \ket{\phi} = \ket{\phi} \ket{0}$. We then perform the desired single-qubit unitary, $u \in U(2)$, \emph{on the ancilla}, giving the state $ u\ket{\phi} \otimes\ket{0}$. We then repeat the qubit-ancilla interaction giving the final state $K [u\ket{\phi} \otimes\ket{0}]=\ket{0} \otimes u\ket{\phi}$. We have performed the unitary $u$ on the register qubit by only manipulating the ancilla and using a fixed interaction twice. This procedure is shown in figure~\ref{singlequbitfig}.
\begin{figure}[h!]
\center
\includegraphics[scale=1.2]{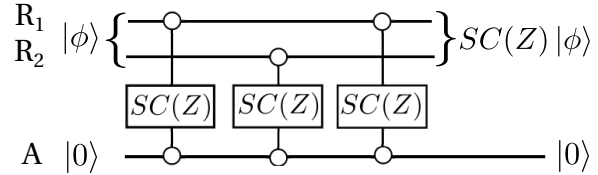}
\caption{\label{twoqubitfig} A simple example of a two-qubit entangling gate in ACQC where the ancilla-register interaction is $SC(Z)$ and the effective register qubit interaction created is also of this form.}
\end{figure}
 It can be seen that if the state of either input qubit is $\ket{0}$ then $K$ has the same action as SWAP. Using this, it is simple to see how we can perform two-qubit gates. Our initial state is now $\ket{0} \otimes \ket{\phi}_{R_{1,2}}$ where $\ket{\phi}$ is the (arbitrary) state of the two register qubits, $R_1$ and $R_2$. We first interact the ancilla with the first register qubit, which swaps the state of that qubit with the ancilla. We then interact the ancilla with the second qubit and finally interact the ancilla with the first qubit again, swapping the $\ket{0}$ state back to the ancilla. This can be seen to be the operation
\begin{equation} K_{AR_1} K_{AR_2} K_{AR_1} \ket{0}_A \ket{\phi}_{R_{1,2}} = \ket{0}_A \otimes K\ket{\phi}_{R_{1,2}}\end{equation}
where identity operators are omitted for simplicity. This is shown in figure~\ref{twoqubitfig}. The overall effect of this sequence of operations is the entangling gate $K=SC(Z)$ between the two register qubits, where a two-qubit gate is called entangling if it can create entanglement between two qubits initially in a separable state \cite{Bremner02}. As a single entangling gate combined with arbitrary single-qubit unitaries is sufficient for universal quantum computation \cite{Bremner02,Brylinski02}, the above scheme is a universal quantum computer, where the only interaction with the register qubits is via the fixed interaction $SC(Z)$. To perform the final measurement readout of a register qubit, that qubit interacts with the ancilla which is then measured.

\section{Interactions for ACQC}
We now characterise all possible forms that $K$ can take for which this form of computation is possible. 
Two operators, $K_{AR}$ and $K^{'}_{AR}$ $\in$ U(4), are called \emph{locally equivalent} if 
\begin{equation} K_{AR}=k_A^{(1)} \otimes k_R^{(1)} \cdot K^{'}_{AR} \cdot k_A^{(2)} \otimes k_R^{(2)} \end{equation}
for some $k_j^{(i)} \in U(2) $.   
\newline
\indent
\emph{Proposition 1a:} The \emph{fixed} ancilla-register interactions, $K_{AR} \in U(4)$, that are sufficient for ACQC must necessarily be locally equivalent to $ SC(p)$ for some $p \in U(2)$, where $C(p)$ is entangling. 
\newline
\indent
The full proof of this is included in appendix A, however it can be outlined as follows. Clearly to implement two-register-qubit entangling gates the ancilla-register interaction must be entangling. We now consider what constraints the need to implement single-qubit unitaries enforces on the form of the interaction.
To implement an arbitrary unitary, $u$, on the register in ACQC, we require that for each $u$ there exists some  $\tilde{u}$ such that  
\begin{equation} K \cdot \tilde{u} \otimes I \cdot K \ket{\psi_0} \ket{\phi} =\ket{\psi_f} \otimes u \ket{\phi} \label{proof} \end{equation} for all $\ket{\phi}$ and some initial ancilla state $\ket{\psi_0}$ which should not depend on $\ket{\phi}$. This is because in this model we require that any single-qubit unitary can be performed by an interaction with the ancilla, an operation on the ancilla and then a second interaction with the ancilla.
If the state $K \ket{\psi_0} \ket{\phi}$ is non-separable for at least some register input, then it can be shown that the final state for that input is non-separable for some $u$, and so hence it is not possible to implement the single qubit unitaries with this choice of interaction and initial ancilla state. This is done by representing $K \ket{\psi_0} \ket{\phi}$ as an entangled state in its Schmidt basis and then showing that the \emph{concurrence} \cite{Wootters98} (a measure of entanglement that is zero for product states) of the final state cannot be zero for every $u$. Hence, as non-zero concurrence implies entanglement, we see that if $K \ket{\psi_0} \ket{\phi}$ is not separable for some $\ket{\phi}$ then we cannot perform the single qubit unitaries and so hence $K \ket{\psi_0} \ket{\phi}$ must be separable for all register input states $\ket{\phi}$. This means that under the ancilla input state $\ket{\psi_0}$, the operator $K$ is equivalent to a non-entangling operator. 
\newline
\indent
Non-entangling operators are locally equivalent to either the identity or SWAP \cite{Bremner02}. For $u$ to take all possible values in $U(2)$ then so must $\tilde{u}$, as the $K$ operator is fixed and so the only effect it can have on $\tilde{u}$ is constant rotations in $U(2)$.  If $K$ is equivalent to a separable operator on the input of the ancilla state $\ket{\psi_0}$ then we may write $K \ket{\psi_0}\ket{\phi}= f\ket{\psi_0}g\ket{\phi}$ for some $f,g \in U(2)$. Under this condition the left hand side of (\ref{proof}) can be written as $ K \cdot \tilde{u}f  \ket{\psi_0} \otimes g\ket{\phi}$. This state cannot be separable for all $\tilde{u}$ and $\ket{\phi}$ as $K$ is entangling and so it cannot create separable states from all separable inputs which implies that, for this form of $K$, (\ref{proof}) cannot hold for all $\tilde{u}$ and $\ket{\phi}$. Therefore $K$ is locally equivalent to SWAP on the ancilla input of $\ket{\psi_0}$. It is then straightforward to show that $K$ must be locally equivalent to $SC(p)$, which concludes the proof.
\newline
\indent
While it is necessary for the interaction to be locally equivalent to $SC(p)$ to implement ACQC, not every gate that is locally equivalent to $SC(p)$ is sufficient. We can instead state the following stronger proposition:
\newline
\indent
\emph{Proposition 1b:}
Let $k_A^{(1)}, k_R^{(1)}, k_A^{(2)}, k_R^{(2)} \in U(2)$ and $\ket{\psi} \equiv k_A^{(2)} \ket{\psi_0}$, where $\ket{\psi_0}$ is the state the ancilla is initialised to before each gate, and define $\ket{\psi_{\perp}}$ such that $\expect{\psi,\psi_{\perp}}=0$. An interaction, $K$, is sufficient for ACQC if and only if $K$ is of the form 
\begin{equation} K =k_A^{(1)} \otimes k_R^{(1)} \cdot SC_{\psi_{\perp}}(p)  \cdot k_A^{(2)} \otimes k_R^{(2)}, \label{p1b} \end{equation}
where $C_{\psi_{\perp}}(p)$ is the operator that implements $p$ on the target (register) qubit if the control (ancilla) qubit is in the state $\ket{\psi_{\perp}}$, for some local unitaries $k^{(i)}_j$ such that $k_R^{(2)}k_R^{(1)}k^{(2)}_A\ket{\psi_0}$ is an eigenstate of $p$. These operators are a subset of those that are locally equivalent to $SC(p)$. 
Given an interaction which can be written in the form of (\ref{p1b}), there is not necessarily an initial ancilla state that satisfies the above conditions. The non-local part of the interaction, $SC_{\psi_{\perp}}(p)$ fixes $\ket{\psi}$. The choice of initial ancilla state is then fixed by  $ \ket{\psi_0}=k_A^{(2) \dagger} \ket{\psi} $ but this will not always satisfy the condition that $k_R^{(2)}k_R^{(1)}k^{(2)}_A\ket{\psi_0}$ is an eigenstate of $p$. Hence it is not only the non-local part of the interaction which determines whether or not it is suitable for ACQC but also the local unitaries. The proof of proposition 1b is shown in appendix A, and follows in a straightforward manner from the proof of proposition 1a.

\section{Hamiltonians for ACQC}
 We will now characterise the possible Hamiltonians capable of this form of computation.  Let $V \in U(4)$, and define $m(V)$ by
\begin{equation} m(V) \equiv (Q^{\dagger} V Q)^{T} Q^{\dagger}V Q, \label{m}\end{equation}
where in the computational basis
\begin{equation} Q = \frac{1}{\sqrt{2}}\begin{pmatrix} 1 & 0 & 0 & i \\ 0 & i & 1 & 0 \\ 0 & i & -1 & 0 \\ 1 & 0 & 0 & -i \end{pmatrix}. \end{equation}
  The \emph{local invariants} of $V$, first introduced by Makhlin \cite{Makhlin02}, are defined as
\[G_1(V) \equiv \frac{\text{tr}^2[m(V)]}{16 \det V } , \]
\begin{equation} G_2(V) \equiv \frac{\text{tr}^2[m(V)]- \text{tr}[m^2(V)]}{4 \det V } \label{1G}, \end{equation}
and are equal for $V$ and $V^{'}$ if and only if $V$ and $V^{'}$ are locally equivalent \cite{Makhlin02,Zhang03}. Using ($\ref{m}$) it is straightforward to show that $m(SV)=m(V)$. Now as $\det[\text{SWAP}]=-1$ we have that
 \begin{equation}\det[SV]=\det[\text{SWAP}]\det[V]=-\det[V].\end{equation}
 From (\ref{1G}) it can then be seen that 
\begin{equation} G_1(SV)=-G_1(V), \hspace{0.5cm} G_2(SV)=-G_2(V) \label{GSV}. \end{equation} We can parameterise a general $U(2)$ matrix, $p$, by
\begin{equation}  p = e^{i \eta}\begin{pmatrix} e^{i \phi}\cos \theta & e^{-i \psi} \sin \theta  \\  e^{i \psi}\sin \theta   & - e^{-i \phi} \cos \theta \end{pmatrix}. \end{equation} 
Using this parameterisation it can be shown that
\begin{equation} G_1(C(p))=\cos^2\theta \sin^2 \phi, \hspace{0.3cm} G_2(C(p))=1+2G_1 \label{R}. \end{equation}
We note that taking $\theta=0$ in $C(p)$ gives a local equivalent of $C(R(-2\phi))$, where $R(\phi ) \equiv \ket{0} \bra{0} + e^{i \phi} \ket{1} \bra{1} $, and as from (\ref{R}) the range of the local invariants for $C(R(-2\phi))$ and $C(p)$ is the same, that each controlled unitary is locally equivalent to a controlled rotation. The local invariants of $C(Z)$ (and CNOT) are $G_1=0$ and $G_2=1$ \cite{Makhlin02}. From (\ref{GSV}) and (\ref{R}) it can then be seen that the operators capable of ACQC have local invariants
\begin{equation} G_1(SC(p))=-\cos^2\theta \sin^2 \phi, \hspace{0.3cm} G_2(SC(p))=-1+2G_1. \label{invars} \end{equation}
\indent
A general operator in U(4) may be parameterised using the \emph{canonical decomposition} as
\begin{equation} K = k_A^{(1)} \otimes k_R^{(1)} \cdot M \cdot k_A^{(2)} \otimes k_R^{(2)}, \end{equation}
where $M = e^{\frac{i}{2}(\alpha_1 X\otimes X + \alpha_2 Y \otimes Y +\alpha_3 Z \otimes Z   )}$, $k_j^i \in U(2) $ and $\alpha_1, \alpha_2, \alpha_3 \in \mathbb{R}$,  where each $\alpha_i$ has period $\pi$ \cite{Khaneja01,Kraus01}. We therefore see that the non-local properties of an operator can be represented by 3 parameters, which are related to its local invariants by 
\begin{multline} G_1 = \cos^2 \alpha_1 \cos^2 \alpha_2 \cos^2\alpha_3-\sin^2\alpha_1\sin^2\alpha_2\sin^2\alpha_3 \\ + \frac{i}{4} \sin2\alpha_1\sin 2\alpha_2\sin 2\alpha_3, \label{G1} \end{multline}
\begin{multline} G_2 = 4 \cos^2\alpha_1\cos^2\alpha_2\cos^2\alpha_3-4 \sin^2\alpha_1\sin^2\alpha_2\sin^2\alpha_3 \\ - \cos 2\alpha_1\cos 2\alpha_2\cos 2\alpha_3, \label{G2} \end{multline}
as shown in \cite{Zhang03}. 
\newline
\indent
Setting the equations of (\ref{invars}) equal to (\ref{G1}) and (\ref{G2}) results in a pair of simultaneous equations for $\alpha_1$, $\alpha_2$ and $\alpha_3$. The solutions are the possible values in the canonical decomposition for which the non-local properties of the interaction are those that allow for ACQC. It is straightforward to show that the possible solutions are given by $\alpha_1 = (2n+1) \frac{\pi}{2}$ and  $\alpha_2 = (2m+1) \frac{\pi}{2} $, where $n,m \in \mathbb{N}$, with remaining coefficient a solution to $\sin^2 \alpha_3=\cos^2\theta \sin^2 \phi$. We can then see that the Hamiltonian
\begin{equation} H = -\hbar \chi(\alpha_1(n) X \otimes X + \alpha_2(m) Y \otimes Y + \alpha_3(\theta,\phi) Z \otimes Z),\end{equation}
applied for a time $t=\frac{1}{2\chi}$, with any choice of $n$ and $m$, implements a local equivalent of the unitary $SC(p)$ where $p=p(\theta,\phi)$. The  only further constraint is that $\theta$ and $\phi$ must be chosen such that the unitary is not locally equivalent to SWAP, or equivalently that the unitary is entangling. The local invariants of SWAP are \cite{Makhlin02} $G_1(\text{SWAP})=-1$ and $G_2(\text{SWAP})=-3$. Hence, from (\ref{invars}), $\theta$ and $\phi$ may not be solutions to $\cos^2 \theta \sin^2 \phi = 1$. Clearly the $\alpha_i$ can be exchanged as $G_1$ and $G_2$ are symmetric in swapping $\alpha_i$. 
From these conditions it is simple to show that we cannot choose all $\alpha_i$ the same as this results in implementing a local equivalent to SWAP. It is also important to note that the local unitaries implemented by the Hamiltonian need to satisfy the conditions for ACQC given in proposition 1b. The above conditions only guarantees that the non-local part of the interaction is as required.
\newline
\indent
We will now find some specific examples of appropriate Hamiltonians. A \emph{perfect entangler} is a $U(4)$ operator that can produce maximally entangled states from product states. A controlled gate is a perfect entangler only if it is locally equivalent to $C(Z)$ \cite{Zhang03}. Therefore, if we wish the interaction, $K$, to be a perfect entangler then we require that $K$ is locally equivalent to $SC(Z)$. From the local invariants of $C(Z)$ we have that $G_1(SC(Z))=0$ and $G_2(SC(Z))=-1$. It is straightforward to show from (\ref{G1}) and (\ref{G2}) that taking $\alpha_1=(n+1)\frac{\pi}{2}$, $\alpha_2=(2m+1)\frac{\pi}{2}$, $\alpha_3=n\frac{\pi}{2}$ in the canonical decomposition, where $n$ and $m$ are integers, gives local equivalents to $SC(Z)$. In particular, letting $n=m=0$ gives the solution $\alpha_1=\frac{\pi}{2}$, $\alpha_2=\frac{\pi}{2}$, $\alpha_3=0$. Therefore the two-qubit XY exchange Hamiltonian
\begin{equation} H_1 = -\hbar \chi (X \otimes X + Y \otimes Y ),\end{equation}
applied for a time $t=\frac{\pi}{4 \chi}$ is locally equivalent to $SC(Z)$. However, we have seen that local equivalence is not enough to infer that the interaction is capable of implementing ACQC. Using the spectral theorem it can be shown that applying this Hamiltonian for a time $t$ gives the unitary 
\begin{equation}  U_1(t) = \begin{pmatrix} 1 & 0 & 0 & 0 \\ 0 & \cos 2\chi t & i\sin 2\chi t & 0 \\ 0 & i\sin 2\chi t & \cos 2\chi t & 0 \\ 0 & 0 & 0 & 1 \end{pmatrix} .\end{equation} 
When $t=\frac{\pi}{4 \chi}$ we have
\begin{equation}  U^c_1 \equiv U_1\left(\frac{\pi}{4 \chi}\right)=  \begin{pmatrix} 1 & 0 & 0 & 0 \\ 0 & 0 & i & 0 \\ 0 & i & 0 & 0 \\ 0 & 0 & 0 & 1 \end{pmatrix} .\end{equation} 
We therefore have that our interaction term is 
\begin{equation} U_1^c= s \otimes s \cdot SC(Z), \end{equation}
where $s=\ket{0}\bra{0}+i\ket{1}\bra{1}$. For this interaction to be capable of implementing ACQC we require that their exists a choice of initial ancilla state that satisfies the conditions of proposition 1b. If we write our interaction in the form $k^{(1)}_A \otimes k^{(1)}_R \cdot SC_{\psi_{\perp}}(p) \cdot k^{(2)}_A \otimes k^{(2)}_R$ then the two conditions of proposition 1b are that the initial ancilla state, $\ket{\psi_0}$, must be such that $k^{(2)}_R k^{(1)}_Rk^{(2)}_A \ket{\psi_0}$ is an eigenstate of $p$ and $\ket{\psi_0}= k^{(2)\dagger}_A \ket{\psi}$, where $\ket{\psi}$ is the state such that $\expect{\psi,\psi_{\perp}}=0$. As the interaction $U_1^c$, expressed in this form, has $p=Z$, $\ket{\psi}=\ket{0}$, $\ket{\psi_{\perp}}=\ket{1}$, $k^{(1)}_A=k^{(1)}_R=s$ and $k^{(2)}_A=k^{(2)}_R=I$ it is straightforward to see that $\ket{\psi_0}=\ket{0}$ satisfies these conditions. It is then simple to show that
\begin{equation}  U^c_1  \cdot u\otimes I \cdot  U^c_1 \ket{0} \otimes\ket{\phi}=\ket{0}\otimes s \cdot u \cdot  s\ket{\phi}.\end{equation} 
Therefore, in order to implement the unitary $u$ on the register, we implement the unitary $\tilde{u}=s^{\dagger}us^{\dagger}$ on the ancilla. It is then possible to show that the effective two-register-qubit entangling operation is $SC(Z)$, as
\begin{equation}  U_{AR_1}^c  U_{AR_2} ^c  U_{AR_1}^c \ket{0}_A \otimes\ket{\phi}_{R_{1,2}} = \ket{0}_A \otimes SC(Z) \ket{\phi}_{R_{1,2}} ,\end{equation} 
up to an irrelevant global phase.
\section{Universal Gate Sets}
Any two-qubit entangling gate, along with arbitrary single-qubit unitaries, is a universal set. It is however desirable, for error correcting and practicality, to find a finite set that is also universal (up to arbitrary accuracy). As much of the work in developing quantum computation is based around the $C(Z)$ or CNOT primitives, it is useful to see how we can simulate CNOT with our interaction. As the local invariants of $SC(Z)$ are not the same as those of CNOT, it is clear that the minimum number of applications of $SC(Z)$ required to implement a CNOT is two. Using a variation on the results of Bremner et. al. in \cite{Bremner02} it is possible to derive that
\begin{equation} \text{CNOT} = X \otimes X \cdot U^c_1 \cdot H Y \otimes Z \cdot U^c_1 \cdot X  s \otimes H s^{\dagger}  H, \end{equation}
up to an irrelevant global phase, where $H = \frac{1}{\sqrt{2}}(\ket{0}\bra{0}+\ket{0}\bra{1}+ \ket{1}\bra{0} - \ket{1}\bra{1})$ is the Hadamard gate. If we can perform each of these single-qubit operations (or simulate them), multiplied on either side by $s^{\dagger}$, then we can simulate CNOT. However CNOT and all these single-qubit gates are members of the Clifford group and so by the Gottesman$-$Knill theorem we need a further single-qubit operation for universal quantum computation based on CNOT simulation \cite{Gottesman99}. Such an additional gate is the $T$ (or $\frac{\pi}{8}$)-gate where $T=\ket{0}\bra{0}+e^{i\frac{\pi}{4}}\ket{1}\bra{1}$. 
\newline
\indent
If however we wished to implement the computation entirely by simulating CNOT gates, then we could choose an alternative unitary with which it is more convenient to simulate CNOT. Such a unitary does not need to be a perfect entangler and so does not need to be locally equivalent to $SC(Z)$. One such choice arises from taking the Hamiltonian
\begin{equation}  H_2 = -\frac{\hbar \chi}{2} \left(2 X \otimes X + 2 Y \otimes Y + Z \otimes Z \right).  \end{equation} 
This Hamiltonian, applied for a time $t$, implements the unitary
\begin{equation}  U_2( t) = e^{i \frac{\chi t}{2}} \begin{pmatrix}1 & 0 & 0 & 0 \\ 0 &  e^{-i \chi t} \cos 2 \chi t & e^{-i \chi t} i \sin 2 \chi t  & 0\\ 0 & e^{-i \chi t} i \sin 2 \chi t & e^{-i \chi t} \cos 2 \chi t  & 0 \\ 0 & 0 & 0 & 1 \end{pmatrix}, \end{equation} 
which at $t=\frac{\pi}{4 \chi}$ gives
\begin{equation}  U_2^c \equiv U_2\left(\frac{\pi}{4 \chi}\right) = \begin{pmatrix} e^{i \frac{\pi}{8}} & 0 & 0 & 0 \\ 0 & 0 &i e^{-i \frac{\pi}{8}} & 0\\ 0 & i e^{-i \frac{\pi}{8}} & 0  & 0 \\ 0 & 0 & 0 &  e^{i \frac{\pi}{8}}\end{pmatrix}. \end{equation} 
It is easy to see that this unitary fulfills the criteria required for ACQC, with the choice of initial ancilla state $\ket{\psi_0}=\ket{0}$, as it is locally equivalent to $SC(s)$, where the local rotations are diagonal. Now we have that
\begin{equation} U_2^c U_2^c= \begin{pmatrix} e^{i \frac{\pi}{4}} & 0 & 0 & 0 \\ 0 & - e^{-i \frac{\pi}{4}} & 0  & 0\\ 0 & 0 &- e^{-i \frac{\pi}{4}} & 0 \\ 0 & 0 & 0 &  e^{i \frac{\pi}{4}}\end{pmatrix} \end{equation} 
which is locally equivalent to C(Z), and hence CNOT, via the local rotation on each qubit of $R= e^{-i \frac{\pi}{8}} \ket{0} \bra{0} - e^{i\frac{3\pi}{8}} \ket{1}\bra{1}$. The local invariants are given by
\begin{equation} G_1\left(U_2^c\right)=- \frac{1}{2}, \hspace{1cm} G_2\left(U_2^c \right)=- 2.\end{equation} 
\indent
The \emph{entangling power} of a $U(4)$ operator, $K$, is defined as the average entanglement produced by $K$ when acting on product states \cite{Zanardi00,Zanardi01}. Entangling power can be related to the local invariants of an operator by
\begin{equation} e_p(K) = \frac{2}{9}(1-|G_1(K)|).\end{equation} Hence, two operators that have the same $|G_1|$ have the same entangling power \cite{Balakrishnan10}. As $|G_1|\leq 1$, we have that $ 0 \leq e_p \leq \frac{2}{9}$. We therefore see that $e_p(U_2^c)=\frac{1}{2}e_p(CNOT)=\frac{1}{9}$. 
\newline
\indent
We can implement the single-qubit unitaries as before, however there is a more convenient way to implement the CNOT gate, by instead implementing
\begin{multline} U^c_{2_{AR_1}} \left( U^c_{2_{AR_2}} \right)^2 U^c_{2_{AR_1}} \ket{0}_{A} \ket{\phi}_{R_{1,2}} \\=  \ket{0}_{A} Z \otimes s\cdot C(Z) \ket{\phi}_{R_{1,2}}.  \end{multline}
By using the same procedure as earlier for single-qubit gates, we can see that implementing the gate $u \in U(2)$ on the ancilla results in the gate $u \cdot s$ being applied to the register qubit. Therefore as $s=T^2$, $Z=T^4$ and $s^{\dagger}=T^6$ we can simulate $C(Z)$ with only $T$ gates and $SC(s)$ interactions with only a minor $T$ gate overhead. As CNOT is locally equivalent to $C(Z)$ via Hadamard rotations and \{CNOT, H, T\} is a universal, and fault tolerant, gate set \cite{Boykin00}, a simple universal gate set for this model of computation is the $SC(s)$ ancilla-register interaction in conjunction with $H$ and $T$ gates on the ancilla.

\section{Discussion}
We have presented a form of universal quantum computation, ACQC, in which the only access required to the computational qubits is via a fixed interaction with a fully controllable ancilla qubit. In contrast to ancilla-driven quantum computation, the ancillary qubits evolve through unitary dynamics only. This removes the need for accurate ancilla measurements, except for the final readout, which can potentially increase the speed and fidelity of computation. In order to keep all of the dynamics unitary in ACQC, it was necessary to increase the number of interactions between the ancilla and register qubits in comparison to ADQC, from one to two in the case of single-qubit gates, and two to three in the case of two-qubit gates. It is clear that the relevant source of errors for this model will then be inaccurate unitary control of both the ancillary system and the ancilla-register interaction. Standard error correction protocols can be applied to this scheme if the ancilla is reset (or a fresh ancilla used) after the completion of each single or multi-register-qubit gate. The resetting of the ancilla will remove any residual entanglement between the ancilla and the register, due to imperfect gate implementation, and hence prevent correlated errors from propagating through the computation.
\newline
\indent
We have shown that any interaction capable of this form of computation is locally equivalent to the SWAP gate combined with an entangling controlled unitary.
It has been shown that the only interactions that allow (stepwise) deterministic ancilla-driven quantum computation, where only Pauli corrections are required, as in the one-way quantum computer, are locally equivalent to $C(Z)$ or $SC(Z)$, although not all locally equivalent interactions are sufficient \cite{Kashefi09}. This restricts the applicability of deterministic ADQC. It can, however, be generalised to further interactions if probabilistic schemes are employed \cite{Shah13}. It is interesting to note that the class of interactions that allow for ACQC and those that allow for the (stepwise) deterministic ancilla driven model overlap but are not the same. In ACQC we are free to use gates of any non-zero entangling power. This is because, unlike in ADQC, the entangling properties of the interaction are not used to implement the single-qubit gates. We have given appropriate Hamiltonians for implementing this model and some simple universal gate sets. In particular, the ability to implement a certain case of the XXZ two-qubit Hamiltonian as the ancilla-register interaction, along with $T$ and $H$ gates on the ancilla, form a simple universal fault tolerant set.
\newline
\indent
The work of TJP was supported by a University of Leeds Research Scholarship.

\appendix
\section{}
In this appendix we prove propositions 1a and 1b.
We wish to show that the only ancilla-register interactions that allows for ACQC are a subset of those locally equivalent to $SC(p)$ where $C(p)$ is entangling, and that all members of this subset are sufficient for ACQC. We will do this by considering the constraints that the need to implement single-qubit gates enforces on the form of this interaction. 
\newline
\indent
When defining the ACQC model of computation, it was stated that we wish \emph{any} single-qubit unitary, $u \in U(2),$ to be implementable on a register qubit by interacting the ancilla with that register qubit twice with ancilla control in between. This was chosen as a condition as, although in practice only a finite universal gate set will be implemented on the ancilla, we want the interactions characterised to be valid for ACQC independent of what particular universal set can be implemented on the ancilla. We now state the precise condition on the interaction. We do this by first stating a relation that holds only for those $K$ that are capable of ACQC, and then justify why interactions for ACQC must satisfy this relation.
\newline
\indent
Take $k_A^{(1)}, k_R^{(1)}, k_A^{(2)}, k_R^{(2)} \in U(2)$ and let $\ket{\psi} \equiv k_A^{(2)} \ket{\psi_0}$, where $\ket{\psi_0}$ is the state to which the ancilla is initialised before each gate. Define $\ket{\psi_{\perp}}$ such that $\expect{\psi,\psi_{\perp}}=0$ and let $K \in U(4)$ be an entangling operator. There exists a $\tilde{u} \in U(2)$ for each $u \in U(2)$ such that
\begin{equation} K \cdot \tilde{u} \otimes I \cdot K \ket{\psi_0}\otimes \ket{\phi} = \ket{\psi_f} \otimes u \ket{\phi} \label{singlequbit} \end{equation} 
$\forall$ $\ket{\phi} \in   \mathbb{C}^{2} $ only if $K =k_A^{(1)} \otimes k_R^{(1)} \cdot SC_{\psi_{\perp}}(p)  \cdot k_A^{(2)} \otimes k_R^{(2)}$ and $k_{R}^{(2)} k_{R}^{(1)} \ket{\psi}$ is an eigenstate of $p$. It is necessary for (\ref{singlequbit}) to hold, for all $u \in U(2)$, in order for the single-qubit unitaries to be implementable in the ACQC model and hence we require that $K$ is of a form that satisfies this condition. The condition that $K$ must be entangling is required for the implementation of the two-qubit gates. This is a restating of proposition 1b. We will now prove the above statement and so hence propositions 1a and 1b.
\newline
\indent \emph{Proof:}
It is necessary for (\ref{singlequbit}) to hold for all possible register input states $\ket{\phi}$. We consider two distinct cases. Either $K \ket{\psi_0} \ket{\phi}$ is separable for all register inputs $\ket{\phi}$, or it is not separable for at least some $\ket{\phi} \in \mathbb{C}^2$.
\newline
\indent
Let us first consider the case where it is not separable for at least some particular register input $\ket{\phi_e}$. Using the Schmidt decomposition theorem, we have that $K \ket{\psi_0} \ket{\phi_e}= \alpha \ket{u_1}\ket{v_1} + \beta \ket{u_2} \ket{v_2}$
for some orthonormal basis' $\{ \ket{u_1}, \ket{u_2} \}$ and $\{ \ket{v_1}, \ket{v_2} \}$, and real, positive coefficients $\alpha$ and $\beta$. As the state is not separable, we have that $\alpha , \beta \neq 0$. 
 Now for (\ref{singlequbit}) to be satisfied for all $u \in U(2)$ we require that
$   K \cdot \tilde{u} \otimes I \cdot K \ket{\psi_0} \ket{\phi_e} $
is separable for all $\tilde{u} \in U(2)$. This is because for $u$ to take all possible values in $U(2)$ then so must $\tilde{u}$, as the $K$ operators are fixed and so the only effect they can have on $\tilde{u}$ is constant rotations in $U(2)$.
\newline
\indent
We parameterise $K$ using the canonical decomposition, giving the condition that $ k_A^{(1)} \otimes k_R^{(1)} \cdot M  \cdot k_A^{(2)} \tilde{u} \otimes k_R^{(2)} \cdot K \ket{\psi_0} \ket{\phi_e}$
is separable for all $\tilde{u}$, where $M = e^{\frac{i}{2}(\alpha_1 X\otimes X + \alpha_2 Y \otimes Y +\alpha_3 Z \otimes Z   )}$. As a Schmidt basis is separable, then if we represent this in a Schmidt basis, $M$ may still be written in the same form as when represented in the computational basis, as the basis changing rotations can be absorbed into the arbitrary $k$ matrices. Using the spectral theorem, in the computational basis
\[ M = \begin{pmatrix} m_1 & 0 & 0 & m_2 \\ 0 & m_3 & m_4 & 0 \\ 0 & m_4 & m_3 & 0 \\ m_2 & 0 & 0 & m_1 \end{pmatrix} \]
with $m_1=e^{i\frac{\alpha_3}{2}} \cos \alpha^-$, $m_2=e^{i\frac{\alpha_3}{2}} i \sin \alpha^-$, $m_3=e^{-i\frac{\alpha_3}{2}} \cos \alpha^+$, $m_4=e^{-i\frac{\alpha_3}{2}} i \sin \alpha^+$
and $\alpha^-=\frac{\alpha_1-\alpha_2}{2}$, $\alpha^+=\frac{\alpha_1+\alpha_2}{2}$.
 We therefore have that in the Schmidt basis of $K \ket{\psi_0} \ket{\phi_e}$ we can write the left hand side of (\ref{singlequbit}) as
 \[ k_{A}^{(1)} \otimes k_R^{(1)} \cdot \begin{pmatrix} m_1 & 0 & 0 & m_2 \\ 0 & m_3 & m_4 & 0 \\ 0 & m_4 & m_3 & 0 \\ m_2 & 0 & 0 & m_1 \end{pmatrix} \cdot k_A^{(2)} \tilde{u} \otimes k_R^{(2)}\begin{pmatrix} \alpha \\ 0 \\ 0 \\ \beta \end{pmatrix} \]
For the left hand side to be equal to the right hand side of (\ref{singlequbit}) we require that the left hand side is separable. Clearly the terms acting after $M$ can have no effect on the separability of the state, and so we require that, relabelling $k^{(2)}_A \tilde{u}$ by $v$, and for labelling simplicity letting $k=k^{(2)}_{R}$ that
 \[ \begin{pmatrix} m_1 & 0 & 0 & m_2 \\ 0 & m_3 & m_4 & 0 \\ 0 & m_4 & m_3 & 0 \\ m_2 & 0 & 0 & m_1 \end{pmatrix} \cdot v \otimes k \begin{pmatrix} \alpha \\ 0 \\ 0 \\ \beta \end{pmatrix} \]
 is separable for all $v \in U(2)$. 
\newline
\indent
We now calculate the concurrence of this state, where the concurrence of a pure state is given by
$ C(\psi)=2 |\psi_{11} \psi_{22} - \psi_{12}\psi_{21}|$ with $\ket{\psi}=\psi_{11}\ket{u_1}\ket{v_1}+\psi_{12}\ket{u_1}\ket{v_2}+\psi_{21}\ket{u_2}\ket{v_1}+\psi_{22}\ket{u_2}\ket{v_2}$. In order for this state to be separable for all $v$ this must be zero for all $v$. By parameterising $v$, this results in 10 simultaneous equations. Here we write only 3 of them as that is all that is required for a contradiction.  They are
\begin{equation} k_{11}^2 m_1m_2 - k_{21}^2m_3m_4=0  \label{sim1}\end{equation}
\begin{equation} k_{12}k_{22} [(m_1^2+m_2^2)-(m_3^2+m_4^2)] = 0 \label{sim2}\end{equation}
\begin{equation} k_{11}k_{22} (m_1^2+m_2^2)- k_{12}k_{21} (m_3^2+m_4^2)=0  \label{sim3}\end{equation}
where the $k_{ij}$ are the components of $k$.
Now if $k_{11}=0$, then due to unitarity we have $k_{22}=0$ and $k_{12}$ and $k_{21}$ are non-zero. From equation (\ref{sim1}), we therefore have that $m_3m_4=0$. But now (\ref{sim3}) gives that $m_3^2+m_4^2=0$. This implies that $m_3=m_4=0$ which contradicts the unitarity of $M$. The same argument applies if $k_{12}=k_{21}=0$. We therefore have that all the components of $k$ are non-zero. We now take a parametrisation of $k$ as
\[ k= e^{i \xi } \begin{pmatrix}e^{i \zeta}\cos \epsilon  &  e^{- i \gamma}\sin \epsilon \\ e^{i \gamma} \sin \epsilon & - e^{-i \zeta}\cos \epsilon  \end{pmatrix} \]
As all the components are non-zero (\ref{sim2}) gives that $ m_1^2 + m_2^2 = m_3^2 + m_4^2 $,
but (\ref{sim3}) gives that
$ (m_1^2+m_2^2) = \frac{k_{12}k_{21}}{k_{11}k_{22}} (m_3^2+m_4^2) $. Therefore $\frac{k_{12}k_{21}}{k_{11}k_{22}} = 1$,
but from our parameterisation of $k$ we have
$\frac{k_{12}k_{21}}{k_{11}k_{22}} = -\frac {\sin^2 \epsilon}{ \cos^2 \epsilon} \leq 0$,
so we have a contradiction. We assumed that $K\ket{\psi_0}\ket{\phi}$ was non-separable for at least some $\ket{\phi}$ and have shown that for that register input the left hand side of (\ref{singlequbit}) is not separable for at least some $\tilde{u} \in U(2)$ and so hence (\ref{singlequbit}) cannot hold.
Therefore, we have shown that the state $K\ket{\psi_0}_A \ket{\phi}_R$ must be separable for all register inputs $\ket{\phi}$ to satisfy (\ref{singlequbit}) for all $\tilde{u}$. 
\newline
\indent
If it is separable for all $\ket{\phi}$, we may write
$ K\ket{\psi_0} \ket{\phi} = \ket{\nu} \ket{\mu} $
for some $\ket{\nu}$ and $\ket{\mu}$. We need this to hold for all register inputs $\ket{\phi}$. As this is a non-entangling mapping for all register inputs, we have that under the ancilla input of $\ket{\psi_0}$, the entangling operator $K$ creates no entanglement for any register input. The only U(4) operators that create no entanglement for all input states are locally equivalent to the identity or to SWAP. Therefore under the ancilla input state $\ket{\psi_0}$, $K$ has the effective operation of one of these two operators. That implies that either $\ket{\nu}=f\ket{\psi_0}$ and $\ket{\mu}=g\ket{\phi}$, or $\ket{\nu}=f\ket{\phi}$ and $\ket{\mu}=g\ket{\psi_0}$ for some $f,g \in$ U(2). Now we require that $  K \cdot \tilde{u} \otimes I \cdot \ket{\nu} \ket{\mu} $ is separable for all $\tilde{u}$. However if we take the first possible choice for $\ket{\nu}$, and $\ket{\mu}$ we have that
 $  K \cdot \tilde{u} \otimes I \cdot f\ket{\psi_0} g\ket{\phi} $ is separable for all $\tilde{u}$  and $\ket{\phi}$. This is not possible as this implies that $K$ creates separable states for all separable input states, which contradicts our demand that $K$ is entangling. We therefore have the latter of our two options, that is that
$ K \ket{\psi_0} \ket{\phi} = f\ket{\phi} g\ket{\psi_0} $
for all $\ket{\phi}$.  We therefore must be able to write $K$ in the form $K=k_{A}^{(1)} \otimes k_{R}^{(1)} \cdot \tilde{K} \cdot k_{A}^{(2)} \otimes k_{R}^{(2)}$ where $k^{(1)}_A k^{(2)}_R=f$, $k^{(1)}_Rk^{(2)}_A=g$ and $\tilde{K} \ket{\psi} \ket{\theta} = \ket{\theta} \ket{\psi} $ for all $\ket{\theta}$, where $\ket{\psi}=k^{(2)}_A\ket{\psi_0}$ and $\ket{\theta}=k^{(2)}_R\ket{\phi}$. In the $\ket{\psi}$, $\ket{\psi _{\perp}}$ basis
\[ \tilde{K} \ket{\psi} (\alpha \ket{\psi} + \beta \ket{\psi _{\perp}})= (\alpha \ket{\psi} + \beta \ket{\psi_{ \perp}})\ket{\psi}  \]
for all $\alpha$, $\beta$ such that $|\alpha|^2+|\beta|^2=1$. This partially defines $\tilde{K}$.
\newline
\indent
If we represent $\tilde{K}$ in the $\ket{\psi}$, $\ket{\psi_{\perp}}$ basis (for both qubits), which we denote $[\tilde{K}]_{\psi}$ to distinguish between an operator and its representation in a particular basis, this fully defines the first two columns as
 \[ [ \tilde{K} ] _{\psi} =  \begin{pmatrix} 1 & 0 & 0 & 0 \\ 0 & 0 & p_{11} & p_{12} \\ 0 & 1 & 0 & 0 \\ 0 & 0 & p_{21} & p_{22} \end{pmatrix} \]
for some unitary $p \in$ U(2). We have that
\[ [ \tilde{K} ] _{\psi} =  \begin{pmatrix} 1 & 0 & 0 & 0 \\ 0 & 0 & 1 & 0 \\ 0 & 1 & 0 & 0 \\ 0 & 0 &0& 1 \end{pmatrix}  \begin{pmatrix} 1 & 0 & 0 & 0 \\ 0 & 1 & 0 & 0\\ 0 & 0 & p_{11} & p_{12}  \\ 0 & 0 & p_{21} & p_{22} \end{pmatrix}.\]
As SWAP has the same matrix representation in any seperable basis' where the same basis is used for each qubit, we therefore have $\tilde{K}=SC_{\psi_{\perp}}(p)$. So we now have that $K=k_{A}^{(1)} \otimes k_{R}^{(1)} \cdot SC_{\psi_{\perp}}(p) \cdot k_{A}^{(2)} \otimes k_{R}^{(2)}$ where $\ket{\psi}=k_{A}^{(2)}\ket{\psi_0}$ and $C_{\psi_{\perp}}(p)$ is entangling. We have therefore proven proposition 1a, as $SC_{\psi_{\perp}}(p)$ is locally equivalent to $SC(p)$. 
\newline
\indent
We now wish to enforce the full required condition of (\ref{singlequbit}), and hence prove proposition 1b. We put the current form of $K$ into this equation which, after applying the first $K$ operator to the state, gives
\[ k_{A}^{(1)} \otimes k_{R}^{(1)} \cdot SC_{\psi_{\perp}}(p) \cdot k_{A}^{(2)} \tilde{u} k_{A}^{(1)}k_{R}^{(2)} \ket{\phi}  \otimes k_{R}^{(2)}k_{R}^{(1)}\ket{\psi} \\ = \ket{\psi_f} u \ket{\phi} \]
Now if the left hand side of this is to be a product state we must have that $C_{\psi_{\perp}}(p) \ket{\eta} \ket{\tau}$ is separable for all $\ket{\eta}$ where $\ket{\eta}=k_{A}^{(2)}\tilde{u} k_{A}^{(1)}k_{R}^{(2)} \ket{\phi} $, $\ket{\tau}=k_{R}^{(2)}k_{R}^{(1)}\ket{\psi}$. Therefore we require $C_{\psi_{\perp}}(p) (\alpha\ket{\psi}+\beta\ket{\psi_{\perp}})  \ket{\tau} =  \alpha\ket{\psi} \ket{\tau} + \beta\ket{\psi_{\perp}} p \ket{\tau}  $ to be separable for all $\alpha$ and $\beta$ such that $|\alpha|^2+|\beta|^2=1$. This is only true if $\ket{\tau}= k_{R}^{(2)}k_{R}^{(1)}\ket{\psi}$ is an eigenstate of $p$. Let the eigenvalue be $e^{i\theta}$. We then have that the final state is
\begin{equation*}k_{A}^{(1)}k_{R}^{(2)}k_{R}^{(1)}k_A^{(2)}\ket{\psi_0} \otimes k_{R}^{(1)}R_{\psi_{\perp}}(\theta)k_{A}^{(2)} \tilde{u} k_{A}^{(1)}k_{R}^{(2)} \ket{\phi} = \ket{\psi_f}u\ket{\phi} \end{equation*} 
where $R_{\psi_{\perp}}(\theta) \equiv \ket{\psi}\bra{\psi} + e^{i \theta} \ket{\psi_{\perp}} \bra{\psi_{\perp}}$. 
So to apply $u$ to the register we apply $\tilde{u}= k^{(2)\dagger}_{A}R_{\psi_{\perp}}(-\theta)k^{(1)\dagger}_{R}uk^{(2)\dagger}_{R} k^{(1) \dagger}_{A} $ to the ancilla. We have shown that $K$ must be of the form $K=k^{(1)}_A \otimes k^{(1)}_R \cdot SC_{\psi_{\perp}} \cdot k^{(2)}_A \otimes k^{(2)}_R$ and furthermore that $\ket{\psi_0}=k^{(2) \dagger}_A \ket{\psi}$ and $k_{R}^{(2)}k_{R}^{(1)}\ket{\psi}$ is an eigenstate of $p$. This is the form of $K$ claimed in proposition 1b.
\newline
\indent
We have seen that all interactions that allow for ACQC must be of this form. It is then simple to see that all such interactions are sufficient for ACQC. The above proof shows how to implement the single-qubit gates. To implement the two-qubit gates we start with the same initial ancilla state as for the single-qubit gates, interact first with one qubit, then with the other qubit and then again with the first. This implements an effective $K$ interaction, up to some local rotations, between the two register qubits. Hence such interactions are necessary and sufficient for ACQC. This concludes the proof of proposition 1b.

\end{document}